# Title: Polarization properties of wurtzite III nitride indicate the principle of polarization engineering

**Authors:**
Kaikai Liu and Xiaohang Li[*]

**Affiliations:**
King Abdullah University of Science and Technology (KAUST), Advanced Semiconductor Laboratory, Thuwal 23955-6900, Saudi Arabia

* Corresponding author Email: xiaohang.li@kaust.edu.sa

**Abstract**:
The spontaneous and piezoelectric polarizations of III-nitrides considerably affect the operation of various III-nitride-based devices. We report an ab initio study of the spontaneous polarization (SP) and piezoelectric (PZ) constants of the III-nitride binary and ternary alloys with the hexagonal reference structure. These calculated polarization properties offer us a profound principle for polarization engineering of nitride semiconductor devices, based on which we propose a few heterojunctions which have nearly-zero polarization effect at the junctions that can potentially enhance optical and power device performances. The polarization doping effect was investigated as well and by BAlN grading from AlN the polarization doping effect can be doubled.

**Main Text:**

Wurtzite (WZ) III-nitride semiconductors and their alloys are widely favorable for optoelectronic devices such as visible and ultraviolet light emitting diodes (LEDs) and laser diodes as well as high-power devices such as high electron mobility transistors (HEMTs). It is both theoretically and experimentally shown that the polarization properties of WZ nitrides have a substantial influence on all the related devices. Because of the asymmetry of the WZ structure, the III-nitrides and their heterojunctions can exhibit strong spontaneous polarization (SP) and piezoelectric (PZ) polarization, which can impact the device operation considerably. For instance, the quantum-confined Stark effect (QCSE) caused by the internal polarization field in the quantum well (QW) of light emitters can reduce radiative recombination rates and shift emission wavelength.[1] In addition, polarization differences at the interface can lead to strong carrier confinement and the formation of two-dimensional electron gas (2DEG), which enables the operation of high electron mobility transistor (HEMT) and impacts the electron blocking layer (EBL).[2,3,4]

Recently, Dreyer et al. showed that the zinc blende (ZB) structure researchers conventionally used as the reference structure needs a significant correction and proposed that the layered hexagonal (H) structure rather than the zincblende (ZB) structure be the reference crystal structure for the evaluation of the SP polarizations.[5] The polarization properties of all the binary nitrides (BN, AlN, GaN and InN) have been calculated with H reference, with which our recent study is quite consistent.[5,6,7] The SP values with ZB reference were widely used for nitride device simulations and designing in the past. Recently, Park et al. studied the optical performance of the B-containing QW structure and reported a significant increase in the UV spontaneous emission rate compared to the conventional AlGaN/GaN QW structure, in which they used both the SP of BN with H reference calculated by Dreyer et al. and the SP values of AlN and GaN with ZB reference.[8,9] This discrepancy of different reference structures makes the polarization properties of III nitrides and their alloys very worthwhile for further device studies. In our previous work, the SP with H reverence and PZ constants of BAlN and BGaN have been calculated.[7] However, those values of the conventional nitride alloys are still absent.

In this work, the SP with H reference and PZ constants of AlGaN, InGaN and InAlN alloys were theoretically calculated. And potential implications for our results are discussed in depth by studying the polarization difference at the heterojunction interface of difference combinations of nitride alloys. Finally, the polarization doping effect is studied based on the polarization properties we calculated.

The calculations were carried out by the Vienna ab initio Simulation Package (VASP) software with generalized gradient approximation (GGA) as the exchange-correlation functional. Our studies show that the polarization properties calculated based on GGA are in good agreement with the ones based on local density approximation (LDA) and hybrid functionals such as Heyd, Scueria and Ernzerhof (HSE).[10,11] The ionic potentials were represented by the projector augmented wave method.[12] The calculations were performed on bulk primitive cells with a 6×6×6 Monkhorst-Pack[13] $k$-point mesh to sample the Brillouin zone at a cutoff energy of 520 eV for the plane-wave basis set. In the structural optimization process, the primitive cells were fully relaxed with the Hellman-Feynman force less than 0.02 eVÅ$^{-1}$. For the alloy calculations, we adopted the 16-atom supercells of the chalchopyritelike (CH) structure to represent 50% alloys and those of the luzonitelike (LZ) structure to represent 25% or 75 % alloys, which was employed and discussed in our last study and other previous works.[14] The SP values with H reference were evaluated using the Berry phase approach[15,16] and the density functional perturbation theory (DFPT) was used to calculate the PZ constants.[17] The lattice constants of the optimized structures of AlGaN, InGaN and InAlN were obtained with small bowing parameters (AlGaN: $b_a = 0.016$ and $b_c = -0.057$; InGaN: $b_a = 0.012$ and $b_c = -0.044$; InAlN: $b_a = 0.053$ and $b_c = -0.136$) and shown in Fig. 1, which is consistent with the previous study.[14]

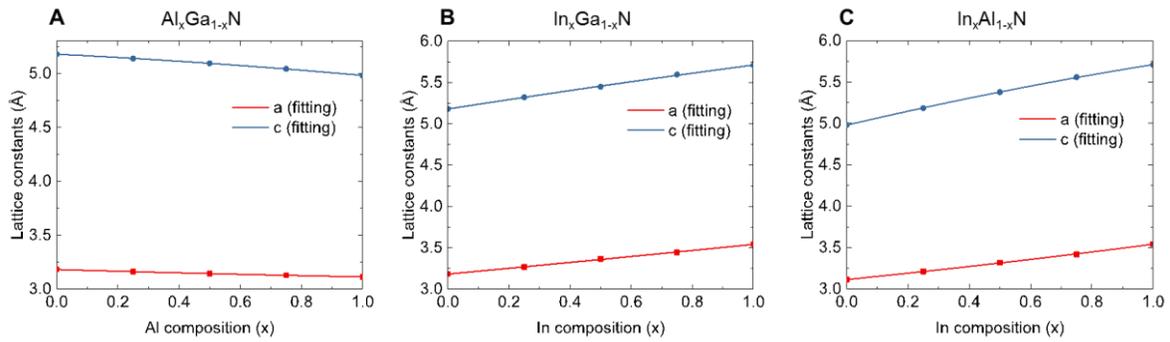

**Fig. 1. Lattice constants of the different alloys.** (**A**) those of AlGaN, (**B**) those of InGaN and (**C**) those of InAlN.

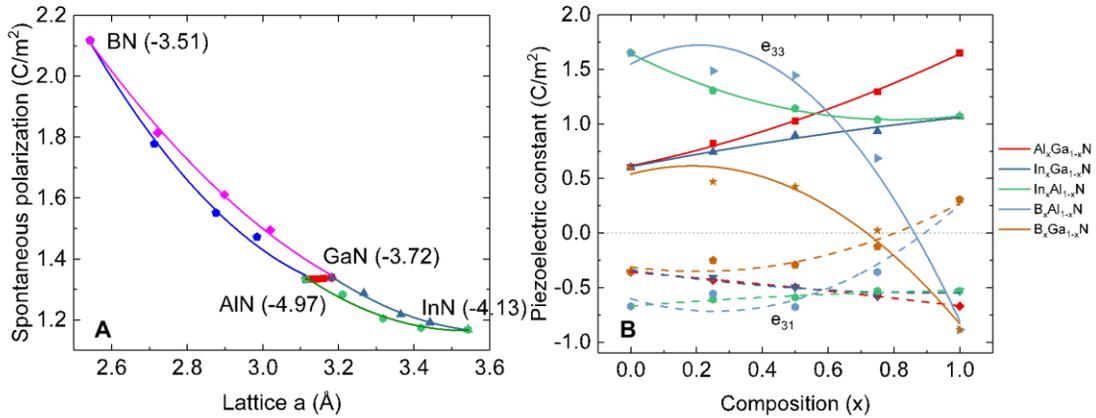

**Fig. 2. Polarization properties of nitride alloys.** (**A**) SP versus the lattice constant $a$. Within the bracket is the effective PZ coefficient $e_{eff}$ in the unit of C/m$^2$. (**B**) PZ constants of AlGaN, InGaN and InAlN.

The SP values of AlGaN, InGaN and InAlN were calculated with H reference in the same way we have calculated the SP values of BAlN and BGaN in our previous study, except that the SP values of InN was calculated by HSE rather than GGA because GGA predicts InN is a conductor. By second-order polynomial regression, we have the following equtions of the SP values of AlGan, InGaN and InAlN,

$$P_{sp}^{(H\ Ref)}(Al_xGa_{1-x}N) = 0.0072x^2 - 0.0127x + 1.3389, \quad (1)$$

$$P_{sp}^{(H\ Ref)}(In_xGa_{1-x}N) = 0.1142x^2 - 0.2892x + 1.3424, \quad (2)$$

$$P_{sp}^{(H\ Ref)}(In_xAl_{1-x}N) = 0.1563x^2 - 0.3323x + 1.3402. \quad (3)$$

Together with the SP values of BAlN and BGaN, we plotted the SP values of all the III nitrides and their alloys versus the lattice constant $a$ in Fig. 2 (A). It is obvious and surprising that regardless of the species of nitride materials the SP value decreases as the lattice becomes larger (except AlGaN), which is consistent with our analysis in our previous study about the physical origin of the nonlinearity of the SP values that is mainly due to the cell volume deformation or the cell volume dilution effect.[7] The PZ constants of AlGaN, InGaN and InAlN are plotted in Fig. 2 (B) and fitted by second-order regression,

$$e_{33}(Al_xGa_{1-x}N) = 0.3949x^2 + 0.6324x + 0.6149, \quad (4)$$

$$e_{31}(Al_xGa_{1-x}N) = -0.0573x^2 - 0.2536x - 0.3582, \quad (5)$$

$$e_{33}(In_xGa_{1-x}N) = -0.1402x^2 + 0.5902x + 0.6080, \quad (6)$$

$$e_{31}(In_xGa_{1-x}N) = 0.2396x^2 - 0.4483x - 0.3399, \quad (7)$$

$$e_{33}(In_xAl_{1-x}N) = 0.9329x^2 - 1.5036x + 1.6443, \quad (8)$$

$$e_{31}(In_xAl_{1-x}N) = -0.0959x^2 + 0.239x - 0.6699. \quad (9)$$

With the SP and PZ constants, we can evaluate the total polarization by the following equations,[5,7]

$$P_{total} = P_{SP} + e_{eff}\epsilon_1, \quad (10)$$

where we define the effective PZ coefficient $e_{eff}$,

$$e_{eff} = 2(e_{31} - P_{SP} - \frac{C_{13}}{C_{33}} e_{33}),  \quad (11)$$

and $\epsilon_1 = \frac{a^{(sub)}(y) - a^{(epi)}(x)}{a^{(epi)}(x)}$ is the strain on the $c$-plane and $C_{13}$ and $C_{33}$ the elastic constants.

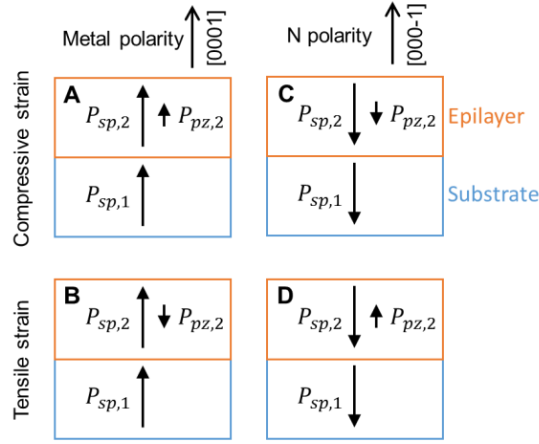

**Fig. 3. SP and PZ polarizations under different situations.** The PZ polarization adds up to the SP value under compressive strain and diminishes the SP value under tensile strain, regardless of metal or N polarity.

**Table 1. Evaluation of the polarization difference of the heterojunctions of various semiconductor devices and proposed heterojunctions with near-zero $\Delta P$.**

| Devices | Heterojunctions | $\Delta P$(C/m$^2$) | Proposed heterojunctions with near-zero $\Delta P$ |
|---|---|---|---|
| HEMT | Al$_x$Ga$_{1-x}$N/AlN ($x$: 0.22 ~ 0.34) | - | Al$_x$Ga$_{1-x}$N/In$_y$Al$_{1-y}$N ($x$: 0.22 ~ 0.34, $y$: 0.7 ~ 0.77) |
| LED laser[18] | In$_{0.15}$Ga$_{0.85}$N/In$_{0.02}$Ga$_{0.98}$N | 0.018 | In$_{0.15}$Ga$_{0.85}$N/In$_{0.48}$Al$_{0.52}$N |
| UVA LED[1] | $Al_{0.1}Ga_{0.9}N/Al_{0.2}Ga_{0.8}N$ | 0.011 | $Al_{0.1}Ga_{0.9}N/In_{0.65}Al_{0.35}N$ |
| Blue LED[19] | $In_{0.2}Ga_{0.8}N/In_{0.02}Ga_{0.98}N$ | 0.026 | $In_{0.2}Ga_{0.8}N/In_{0.51}Al_{0.49}N$ |
| Green LED[20] | $In_{0.25}Ga_{0.75}N/GaN$ | 0.037 | $In_{0.25}Ga_{0.75}N/In_{0.53}Al_{0.47}N$ |

As shown in Fig. 2 (B), unlike BAlN and BGaN alloys whose $e_{33}$ could be negative and $e_{31}$ could be positive,[7] $e_{33}$ and $e_{31}$ of AlGaN, InGaN and InAlN are all positive and negative, respectively. This ensures that $e_{eff}$ of AlGaN, InGaN and InAlN are all negative by evaluating Eq.(11) and specifically those of AlN, GaN and InN are -4.97, -3.72 and -4.13, respectively. The SP value of BN is so large that $e_{ff}$ of BN is negative (-3.51) as well and our careful calculations showed that $e_{eff}$ of BAlN and BGaN are also negative. The fact that all of the III nitrides and their alloys have negative $e_{eff}$ values is still consistent with our conclusion about the nonlinearity of the SP values, which means the total polarization becomes smaller as the cell volume is stretched or the material is under tensile strain in the $c$-plane and vice versa. Therefore, the polarization properties with the SP and PZ constants as a whole indicate that the cell volume is the determinant factor of the polarization value. This has a profound implication for polarization engineering of nitride-based semiconductor devices. For example, if a thin layer with a smaller lattice and a larger SP value is grown on the substrate and fully strained, the tensile strain can act as a volume dilution and potentially leads to zero polarization difference between the epitaxial layer and the substrate due to the induced negative PZ polarization and vice versa, as summarized by Fig. 3.

In the following content, we give several examples to verify this idea by evaluating the polarization difference between different epitaxial layers and different substrate layers, which is given by

$$\Delta P = [P_{sp}^{(epi)}(x) - P_{sp}^{(sub)}(y)] + e_{eff}^{(epi)}(x)\epsilon_1(x,y), \tag{12}$$

where in evaluating $e_{eff}^{(epi)}(x)$ the elastic constants of the epitaxial layer are based on the linear interpolation of the binary values.[10] As shown in Fig. 4, we plotted the polarization difference versus the alloy composition $x$ of the epitaxial layer and the alloy composition $y$ of the substrate of the heterojunction formed by all the combinations of BAlN, BGaN, AlGaN, InGaN and InAlN except the heterojunction of BAlN/BGaN that was reported in our previous study and those formed between B-incorporated and In-incorporated alloys that can have too large lattice mismatch and unrealistic in experiment.

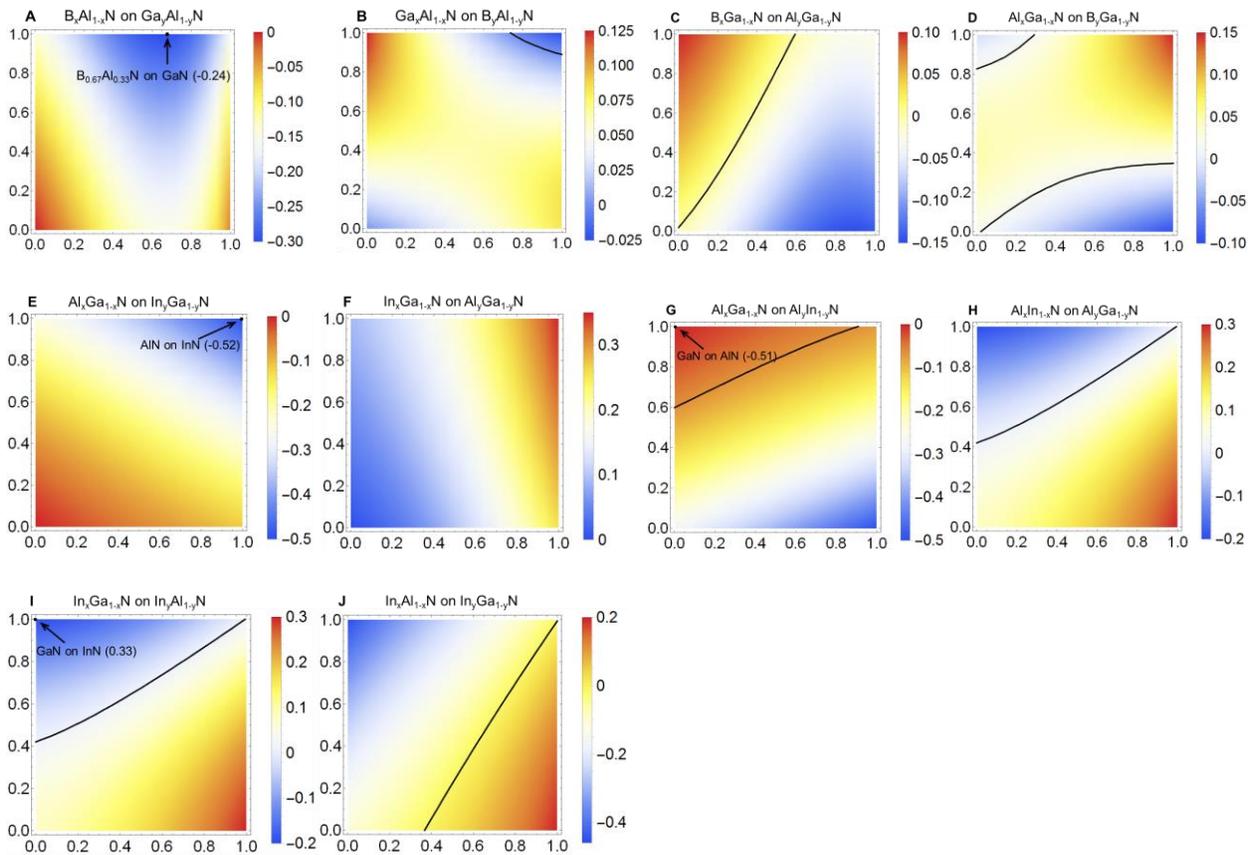

**Fig. 4. Estimated heterointerface polarization difference Δ$P$ (C/m$^2$) between the epitaxial layer and the substrate of metal polarity.** (**A~J**) are the plots of different combinations of the epitaxial layer and the substrate. Black lines indicate zero Δ$P$. Positive Δ$P$ means that along (0001)

direction the epitaxial layer has a larger polarization than the substrate and there is negative charge accumulation at the heterointerface. Large-polarization points are pointed out and marked with different heterojunctions and corresponding $\Delta P$ values in the brackets.

Here we identify two patterns in all the plots in Fig. 4. Since in every plot either at (0, 0) or (1, 1) point $\Delta P$ equals zero because of the same material of the epitaxial and substrate layers and near this point we can approximately model the polarization difference by

$$\Delta P = ax + by \text{ or } \Delta P = a(1-x) + b(1-y). \tag{13}$$

The phenomena that near this point there is either a black line or not depends on $a$ and $b$ in the above equation have either opposite signs or same signs, which are identified as two distinctive patterns and we call pattern 1 and 2. Furthermore, for pattern 1 (BAlN and AlGaN heterojunctions and InGaN and AlGaN heterojunctions), as $x$ and $y$ increase from (0, 0) point or $x$ and $y$ increase from (1, 1) point, one of the lattice constants of the epitaxial and substrate layer increases and the other decreases and they change in opposite directions, which makes the lattice mismatch larger and larger. On the other hand, for pattern 2 (BGaN and AlGaN heterojunctions, InAlN and AlGaN heterojunctions and InGaN and InAlN heterojunctions), as $x$ and $y$ increase from (0, 0) point or $x$ and $y$ increase from (1, 1) point, the lattice constants of the epitaxial and substrate layer change in the same direction, which means that these heterojunctions potentially can have moderate or small lattice mismatches, and meanwhile there exist black lines indicating $\Delta P = 0$. This makes our result very promising for future nitride-based optical device designing. In Table 1, typical haterojunctions in LED lighting devices are listed and the corresponding polarization differences $\Delta P$ are evaluated. For each heterojunction, we proposed another one with near-zero $\Delta P$ based on Fig. 4 that can potentially replace it and offer better device performance.

In addition to estimating the polarization effect at the heterojunctions, we also demonstrate the effectiveness and usefulness of the polarization properties we calculated by considering the polarization doping effect by the ternary alloy grading with more grading options. It was experimentally demonstrated that by grading AlGaN up to ~40% Al composition with a thickness of ~85 nm and in N polarity on top of GaN a p-type AlGaN layer was achieved to have a hole concentration of up to ~$10^{18}$ cm$^{-3}$.[21] Theoretically, if we consider grading the alloy $X_xY_{1-x}N$ fully strained on the substrate of YN of either metal or N polarity, where XN and YN can be those binaries discussed in this work, we can obtain the total polarization of the alloy with any composition by taking the SP and PZ polarization into consideration and the polarization doping concentration is given by the following equations,

$$\sigma_P = -\frac{dP}{dl} = -\frac{dP}{dx}\frac{dx}{dl} \tag{14}$$

$$\kappa = \frac{dP}{dx}, \nu_g = \frac{dx}{dl} \tag{15}$$

where $P$ is the polarization along $c$-axis, $\sigma_P$ is the polarization charge density and $l$ denotes the position of certain grading composition $x$. In the above equations, the polarization doping effect is separated into two independent parts. The first part, $\kappa = \frac{dP}{dx}$, is the polarization changing rate with respect to the alloy composition and can be defined as the composition-polarization changing rate, which only depends on the material itself. The other part, $\nu_g = \frac{dx}{dl}$, can be defined as the grading speed that depends on the experimental aspect. In Fig. 5, we choose GaN or AlN as the substrate as GaN and AlN are the most commonly used substrate for the alloy grading and then we can evaluate composition-polarization changing rate $\kappa$ along six different grading path. Here our evaluation on $\kappa$ is limited only within 30% alloy composition, for the ternary alloy with large grading composition can have large lattice mismatch and practically experimental alloy grading doesn't exceed 40%.[21] Moreover, limiting our evaluation of κ within 30% alloy composition ensures that the polarization effect can be approximately linear. In addition, we only consider the metal-polarity cases, as in the N-polarity cases the direction of the polarization flips over and the signs of all the polarization-related values change their signs with all the laws and properties unchanged.

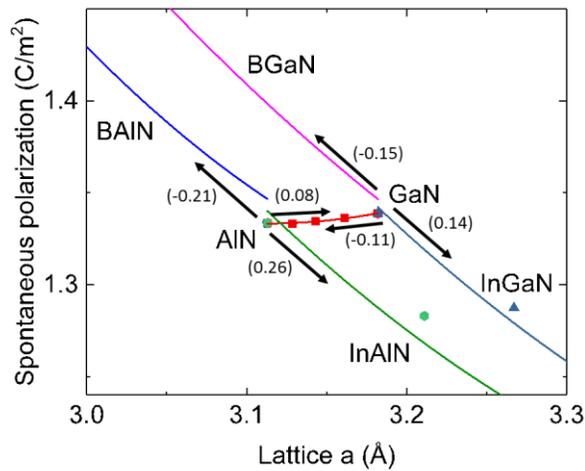

**Fig. 5. Composition-polarization changing rate *κ* (as put in brackets and in the unit of C/m$^2$).** *κ* is evaluated along different grading paths from the starting point of choosing GaN or AlN as the substrate.

From the theoretical calculation, the composition-polarization changing rate *κ* of AlGaN grading on GaN in metal polarity is calculated to be 0.11 C/m$^2$ while according to Simon et al.[21] the experimental *κ* value of that in N polarity was -0.08 C/m$^2$, which are very consistent with each other. Furthermore, here we note three points of interests in Fig. 5. First, the composition-polarization changing rate *κ* of the BAlN grading from AlN is -0.21 C/m$^2$ and that of the InAlN grading from AlN is 0.26 C/m², which are around twice of that of the AlGaN grading from GaN. Second, we notice that in Fig. 5 $\kappa$ is positive along the paths from smaller lattice to larger lattice where the SP values decrease (InAlN grading from AlN and InGaN grading from GaN) or slightly increase (AlGaN grading from AlN) and $\kappa$ is negative along the paths from larger lattice to smaller lattice where the SP values increase (BAlN grading from AlN and BGaN grading from GaN) or slightly decrease (AlGaN grading from GaN), which implies that in polarization doping the strain-induced PZ polarization is more dominant than the spontaneous polarization and the effective PZ coefficients $e_{PZ}$ in Fig. 2 (A) are of equal importance as the SP values. Third, $\sigma_P$ being negative means three dimensional electron gas (3DEG) and $\sigma_P$ being positive means three dimensional hole gas (3DHG), which can be determined by $\kappa$. $\kappa$ can be either negative or positive depending on grading paths and the polarity. For example, Simon et al. achieved 3DEG by AlGaN grading from GaN of N polarity, which means that 3DHG can be achieved by AlGaN grading from GaN of metal polarity, vice versa.

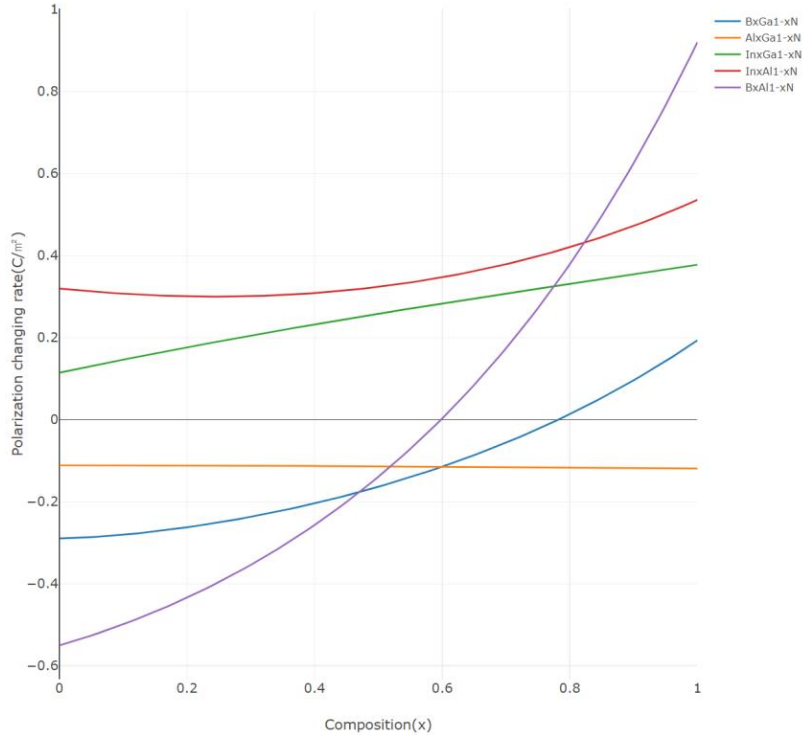

**Fig. 6. Composition-polarization changing rate *κ* of III-nitride ternary alloys, assuming they are strained to the relaxed GaN substrate.**

Moreover, although AlGaN is the most studied alloy among III-nitrides for polarization doping, we have found that it is actually the 'worst' polarization doping candidate due to its small *κ* absolute values in large compositional range compared to the other ternary alloys as shown in Fig. 6. This indicates that other alloys may be utilized to realize more superior polarization doping which has been neglected before. Also AlGaN has the most stable *κ* value which is almost unchanged compared to that of other alloys. Although the results in**Fig. 6** Fig. 6 assume that the alloys are strained to the relaxed GaN substrate, similar results can be obtained assuming that the alloys are partially relaxed to the relaxed GaN substrate or strained to other types of common substrates.

In summary, we calculated the SP and PZ constants of AlGaN, InGaN and InAlN. Together with the SP and PZ constants of BAlN and BGaN, we discovered that different nitride alloys show a consistent relationship between the total polarization and the lattice constant and the total polarization of any nitride alloy can be mainly determined by the lattice constant, which offers us a profound principle for polarization engineering of nitride semiconductor devices. That is, if a thin layer with a smaller lattice and a larger SP value is grown on the substrate and fully strained, the tensile strain can act as a volume dilution and potentially leads to zero polarization difference between the epitaxial layer and the substrate due to induced negative PZ polarization and vice versa. We further investigated the polarization effect of potential heterojunctions to verify this idea and proposed a promising scheme for advancing the performance of nitride optical devices.

Finally, we theoretically study the polarization doping effect and our results reveal that the polarization doping effect of the BAlN grading from AlN or the InAlN grading from AlN is twice of that of the conventional AlGaN grading from GaN.

**Acknowledgments:**

The KAUST authors would like to acknowledge the support of GCC Research Program REP/1/3189-01-01 and KAUST Baseline Fund BAS/1/1664-01-01. We acknowledge the helpful discussions with Feras AlQatari, Haiding Sun, Jingtao Li, and Wenzhe Guo at KAUST.